\documentclass[reprint,superscriptaddress,nofootinbib,amsmath,amssymblrslwwweelqfw,floatfix,float,aps,prd]{revtex4-2}
\usepackage{graphicx}
\usepackage{bm}
\usepackage{amssymb} 
\usepackage[colorlinks=true, allcolors=blue]{hyperref}
\usepackage{cleveref}
\usepackage{xcolor}
\usepackage[normalem]{ulem}

\usepackage{amsmath}
\NewDocumentCommand\mfs{O{}m}{{\color{red} \sout{#1}}~{\color{red} #2}}
\usepackage{placeins}

\begin{document}
\author{Matheus F. S. Alves}
\email{matheus.s.alves@edu.ufes.br}
\affiliation{Departamento de Física \& Núcleo de Astrofísica e Cosmologia (Cosmo-Ufes), Universidade Federal do Espírito Santo, Vitória, ES,  29075-910, Brazil.}
\author{Bruno P. Pônquio}
 \email{bp2799@gmail.com}
\affiliation{Departamento de Física Teórica e Experimental, Universidade Federal do Rio Grande do Norte, Natal, RN, 59078-970, Brazil}
\author{L.G. Medeiros}
\email{leo.medeiros@ufrn.br}
\affiliation{Escola de Ci\^encia e Tecnologia, Universidade Federal do Rio Grande
do Norte, Campus Universit\'ario, s/n\textendash Lagoa Nova, CEP 59078-970,
Natal, Rio Grande do Norte, Brazil.}

\title[]{Critical masses and numerical computation of massive scalar quasinormal modes in Schwarzschild black holes}
\begin{abstract} 
    We present a comprehensive analysis of the quasinormal modes (QNMs) of a massive scalar field in Schwarzschild spacetime using two complementary numerical techniques: the Hill-determinant method and Leaver’s continued-fraction method. Our study systematically compares the performance, convergence, and consistency of the two approaches across a wide range of field masses and angular momenta. We identify three critical mass thresholds, $m_{\rm lim}$, $m_{\rm max}$, and $m_{zd}$, which govern qualitative changes in the QNM spectrum. In particular, long-lived modes emerge at $m_{zd}$, where the imaginary part of the frequency vanishes and the mode becomes essentially nondecaying. This phenomenon is robust across multipoles and may have important implications for the phenomenology of massive fields around black holes. Our results provide a detailed numerical characterization of massive scalar QNMs and highlight the complementary strengths of the Hill-determinant and continued-fraction methods, paving the way for future studies of rotating or charged black holes and quasibound states.

\end{abstract}

\maketitle
\section{INTRODUCTION}
Black holes (BHs) are ideal laboratories for probing the nature of gravity in its most extreme regime. Their response to perturbations is characterized by QNMs, damped oscillations whose frequencies are determined exclusively by the background geometry and the nature of the perturbing field. These oscillations dominate the ringdown stage of astrophysical processes, such as binary mergers, and have become a central element in the interpretation of gravitational wave signals\cite{Vishveshwara:1970zz,chandrasekhar1998mathematical,Kokkotas:1999bd, Nollert:1999ji,Berti:2009kk,Konoplya:2011qq,LIGOScientific:2016aoc,LIGOScientific:2017vwq,EventHorizonTelescope:2019dse,EventHorizonTelescope:2022wkp}.

Perturbations of black holes and the associated QNMs have been extensively investigated. 
Seminal analyses of Schwarzschild stability~\cite{Regge:1957td,Zerilli:1970wzz} were soon extended to 
Reissner-Nordström~\cite{Leaver:1990zz,Kokkotas:1988fm} and Kerr black holes~\cite{Teukolsky:1972my}, 
and later to a variety of compact objects within and beyond general relativity~\cite{Correa:2024xki,Molina:2010fb,
Blazquez-Salcedo:2016enn,Macedo:2018txb,Berti:2018vdi,Antoniou:2024jku,Zinhailo:2024jzt}. 
Beyond their role in gravitational wave astronomy, QNMs are also relevant for astrophysical processes such as 
the dynamics of accretion disks, active galactic nuclei, and gamma-ray bursts. 
Following Unruh’s proposal of black hole evaporation analogs in laboratory systems~\cite{Unruh:1980cg}, 
analog black holes have become a thriving area of research~\cite{Dolan:2011ti,Torres:2020tzs,Barcelo:2005fc,
Matyjasek:2024uwo,dePaula:2025fqt}.

Accurate computation of QNMs requires dedicated numerical techniques, since the perturbation 
equations reduce to wavelike forms with effective potentials that lack closed-form solutions. Several approaches 
have been developed, including direct numerical integration~\cite{Chandrasekhar:1975zza}, WKB-based methods~\cite{1985ApJ...291L..33S,Iyer:1986np,Konoplya:2003ii,Matyjasek:2017psv,Galtsov:1991nwq}, 
the asymptotic iteration method~\cite{Cho:2009cj}, and spectral collocation schemes~\cite{Jansen:2017oag}. 
Among these, two recurrence-based techniques stand out: Leaver’s continued-fraction method~\cite{Leaver:1985ax} 
and the Hill-determinant method~\cite{Majumdar:1989tzg}. The continued-fraction approach is widely regarded as 
the standard, particularly following Nollert’s convergence improvement~\cite{Nollert:1993zz} and subsequent 
refinements~\cite{Zhidenko:2006rs}. The Hill-determinant method, though less common, has recently been shown to achieve comparable accuracy with modern convergence accelerators~\cite{Benda:2025tni}.

In this work, we revisit the QNMs of a massive scalar field in Schwarzschild spacetime with two complementary goals. 
First, we provide a systematic comparison between the Hill-determinant method \cite{Majumdar:1989tzg, Matyjasek:2021xfg} and Leaver’s continued-fraction method \cite{Leaver:1990zz, Leaver:1985ax} in the massive case, assessing their agreement, stability, and numerical efficiency. 
Second, we investigate in detail the impact of the scalar field mass on the QNM spectrum. 
Our analysis reveals the existence of three distinct mass thresholds: $m_{\rm lim}$, $m_{\rm max}$, and $m_{zd}$, which govern qualitative changes in the spectral properties. 
In particular, we show that at $m_{\rm max}$ the quasinormal spectrum undergoes a qualitative change: 
while the classical picture suggests the absence of physical modes for $m > m_{\rm max}$, both methods 
consistently predict the emergence of long-lived modes at $m_{zd}$. These modes, with 
diverging damping times, persist across multipoles and may have important implications for the 
phenomenology of massive fields around black holes.

The paper is organized as follows. In Sec.\ref{sec:massivescalar}, we introduce the perturbation equation for massive scalar fields in Schwarzschild spacetime and outline the Frobenius expansion underlying the Hill-determinant method. Sec.\ref{sec:hilldeterminant} reviews the Hill-determinant approach, while in Sec.\ref{sec:leavermethod} we present the formulation of Leaver's continued-fraction method. In Sec.\ref{sec:results}, we provide a numerical analysis and compare the performance of the two methods. Finally, in Sec.\ref{sec:conclusions} summarizes our main results and discusses possible extensions.

We use the metric signature $(- + + +)$ and $G=c=1$ units.

\section{ MASSIVE SCALAR FIELD IN SCHWARZSCHILD SPACETIME \label{sec:massivescalar}}

The Schwarzschild spacetime is a static and spherically symmetric BH solution.
The line element that describes such spacetime in the standard
Schwarzchild-like coordinates $\left(  t,r,\theta,\phi\right)  $ is%

\begin{equation}
ds^{2}=-f\left(  r\right)  dt^{2}+\frac{1}{f\left(  r\right)  }dr^{2}+r\left(
d\theta^{2}+\sin^{2}\theta d\phi^{2}\right)  , \label{eq:SchLine}%
\end{equation}
where we have defined
\[
f\left(  r\right)  =1-\frac{2M}{r},
\]
with $M$ being the mass of the BH.

Consider now a massive scalar field $\Phi.$ The massive scalar field in a
curved background is governed by the Klein-Gordon equation:%
\begin{equation}
\square\Phi-m^{2}\Phi=\frac{1}{\sqrt{-g}}\partial_{\nu}\left(  g^{\mu\nu}%
\sqrt{-g}\partial_{\mu}\Phi\right)  -m^{2}=0, \label{eq:ScalarEq}%
\end{equation}
where $g$ is the metric determinant, and $g^{\mu\nu}$ the contravariant
metric. Given the spherical symmetry of the spacetime under consideration, it is convenient to decompose the scalar field as follows:
\begin{equation}
    \Phi(t,r,\theta,\varphi) = \frac{\Phi(t,r)}{r}Y_{lm}(\theta,\varphi), \label{separationangles}
\end{equation}
where $Y_{l,m}(\theta,\varphi)$ denote the spherical harmonics with $l$ and $m$ the angular momentum and magnetic numbers, respectively, and $\Phi(t,r)$ is an angle-independent wave function. We can further separate the time dependence by introducing the field’s angular frequency \(\omega\), adopting the so-called harmonic ansatz
\begin{equation}
\Phi(t, r) \equiv \Phi(r) e^{-i \omega t}. \label{separationtime}
\end{equation}
Thus, after substituting Eq. (\ref{eq:SchLine}) into Eq. (\ref{eq:ScalarEq})
using Eq. (\ref{separationangles}) and Eq.(\ref{separationtime}) we obtain:
\begin{equation}
\Phi''\left(  r\right)  
+\frac{1}{r(r-1)}\Phi'\left(  r\right)+\frac{r}{(r-1)^2}\left[\frac{\omega^2}{r}-V(r)\right] \Phi\left(  r\right)  
\label{eq:RadialEq}%
\end{equation}
where
\begin{equation}
V\left(  r\right)  =\frac{(r-1)}{r^4}\left( r^3 m^2-l(l+1)r-1\right) ,
\label{eq:EffectivePot}%
\end{equation}
is the effective potential, with $l=0,1,2,3$... parametrizing the field
angular harmonic index. From here on we use units where $2M=1$. It is important to note that, unlike the usual approach, we are not using tortoise coordinates. As a result, Eq.~(\ref{eq:RadialEq}) does not take the standard Schrödinger-like form.

Physical modes in black hole spacetimes must appear purely ingoing at the
horizon to a local observer, imposing the boundary condition
\begin{equation}
\Phi\left(  \omega,r\right)  \sim e^{-i\omega r}\text{,
}r\rightarrow1.
\end{equation}
At spatial infinity $(r\rightarrow+\infty)$ in an asymptotically flat spacetime, the solution behaves
as%
\begin{equation}
\Phi\left(  \omega,r\right)  \sim A\left(  \omega\right)
e^{-i\sqrt{\omega^{2}-m^{2}}r}+B\left(  \omega\right)  e^{i\sqrt{\omega
^{2}-m^{2}}r}.
\end{equation}
There are two special classes of solutions of physical interest: QNM, characterized by $A\left(  \omega\right)  =0$, corresponding to
purely outgoing waves at spatial infinity and quasibound states (QBS) which are spatially localized and decay
exponentially away from the black hole. In both cases, the imposition of
appropriate boundary conditions at the horizon and at infinity leads to a
discrete spectrum of allowed complex frequencies.

To better understand the differences between QNMs and QBSs, and to properly characterize these solutions, it is essential to analyze their behavior in the asymptotic regions. In this context, techniques for solving Eq.~(\ref{eq:RadialEq}) reveal that the frequencies $\omega$ generally take complex values of the form
\begin{equation}
\omega = \omega_R - i \omega_I, \quad \omega_R, \omega_I \in \mathbb{R}.
\label{def:omega}
\end{equation}

The condition \( \omega_I > 0 \) is necessary to ensure exponential decay at spatial infinity, which is consistent with the stability of the system. Indeed, considering:
\begin{equation}
\Phi \sim e^{-i\omega(t - r)} = e^{-i\omega_R(t - r)} e^{-\omega_I(t - r)},
\end{equation}
we see that \( \omega_I > 0 \) guarantees temporal decay for large \( r \).

Unlike the massless case, Eq.~(\ref{eq:RadialEq}) does not admit only wavelike solutions%
\footnote{By wavelike we mean functions of the form \( f(t \pm r/v) \), with \( v \) the phase velocity.}. 
For real \( \omega^2 \) and \( m^2 \), propagation requires \( \omega^2 > m^2 \). In the present setting, however, 
\( \omega^2 \) is complex, and a more careful analysis is needed. To this end, we define:

\begin{equation}
q^2 = \omega^2 - m^2.
\label{q2}
\end{equation}
Since \( q^2 \), and hence \( q \), are complex, we write:
\begin{equation}
q = q_R - i q_I.
\label{def:q}
\end{equation}

Therefore,
\begin{equation}
\lim_{r \to \infty} \Phi\sim e^{i(q_R - i q_I)r} = e^{i q_R r} e^{q_I r}.
\end{equation}

In this context, the sign of \( q_I \) determines the nature of the solution:
\begin{itemize}
    \item If \( q_I > 0 \), we have a temporal decay \( e^{-\omega_I t} \) and a spatial growth \( e^{q_I r} \). This combination describes a damped perturbation that propagates to infinity — a QNM. Physically, this implies a nonzero energy flux at large distances.
    \item If \( q_I < 0 \), both terms decay, and the perturbation remains spatially localized — a QBS. In this case, the energy flux vanishes at spatial infinity.
\end{itemize}
After characterizing the solutions, we can solve Eq.~(\ref{eq:RadialEq}) by expressing the solution to the wave equation as a Frobenius series. An appropriate ansatz is
\begin{equation}
\Phi\left(  \omega,r\right)  =\left(  \frac{r-1}{r}\right)  ^{\rho
}r^{-\nu}e^{-\nu\left(  r-1\right)  }\sum\limits_{j=0}^{\infty}a_{j}\left(
\frac{r-1}{r}\right)  ^{j}, \label{eq:AnsatzMassive}%
\end{equation}
with $\rho=-i\omega$ and $\nu=-i\sqrt{-\rho^{2}-m^{2}}$. Inserting
Eq. (\ref{eq:AnsatzMassive}) into Eq. (\ref{eq:RadialEq}) we obtain a five-term
recurrent relation for the coefficients $a_{n}$:%
\begin{align}
\alpha_{0}a_{1}+\beta_{0}a_{0}  &  =0,\label{eq:FiveTermRecurrent}\\
\alpha_{1}a_{2}+\beta_{1}a_{1}+a_{0}\gamma_{1}  &  =0,\nonumber\\
\alpha_{2}a_{3}+\beta_{2}a_{2}+a_{1}\gamma_{2}+a_{0}\delta_{2}  &
=0,\nonumber\\
\alpha_{j}a_{j+1}+\beta_{j}a_{j}+a_{j-1}\gamma_{j}+a_{j-2}\delta_{j}%
+a_{j-3}\sigma_{j}  &  =0\text{, \ \ \ }j\geq3,\nonumber
\end{align}
where
\begin{align}
\alpha_j &= (j + 1)(j + 2\rho + 1), \label{eq:CoeffFiveTerm} \\
\beta_j &= -2j - 2\nu - 2\rho - 4j\nu - 8j\rho - 4\rho^2 - 4\nu\rho \nonumber \\
        &\quad -4j^2 - 1 - m^2 - l(l + 1), \\
\gamma_j &= 6j^2 + 10j\nu + 12j\rho - 6j + 4\nu^2 + 10\nu\rho \nonumber \\
         &\quad - 4\nu + 6\rho^2 - 6\rho + 3 + 2l(l + 1), \\
\delta_j &= -4j^2 - 8j\nu - 8j\rho + 10j - 4\nu^2 - 8\nu\rho \nonumber \\
         &\quad + 10\nu - 4\rho^2 + 10\rho - 7 - l(l + 1), \\
\sigma_j &= (j + \nu + \rho - 2)^2.
\end{align}
An important limiting case is the massless limit. From the definitions of
$\rho$ and $\nu$, this corresponds to $m\rightarrow0$ and $\nu\rightarrow\rho
$. Taking the massless limit of Eq. (\ref{eq:CoeffFiveTerm}), we then obtain
\begin{align}
\alpha_{j}  &  =\left(  j+1\right)  \left(  j+2\rho+1\right)
,\label{eq:CoeffFiveTermMassless}\\
\beta_{j}  &  =-4j^{2}-12j\rho-2j-8\rho^{2}-4\rho-1-l\left(  l+1\right)
,\nonumber\\
\gamma_{j}  &  =6j^{2}+22j\rho-6j+20\rho^{2}-10\rho+3+2l\left(  l+1\right)
,\nonumber\\
\delta_{j}  &  =-4j^{2}-16j\rho+10j-16\rho^{2}+20\rho-7-l\left(  l+1\right)
,\nonumber\\
\sigma_{j}  &  =\left(  j+2\rho-2\right)  ^{2}.\nonumber
\end{align}
This yields a massless five-term recurrence relation, which does not allow for
a direct comparison with the three-term relation obtained in Ref.
\cite{Leaver:1985ax}. Nevertheless, in Sec. \ref{sec:results} we show that both formulations reproduce the same quasinormal mode spectrum.

\section{ THE HILL DETERMINANT METHOD\label{sec:hilldeterminant}}

The Hill determinant method, originally developed in the context of periodic
differential equations in mathematical physics, has found significant
application in the computation of QNMs of black holes. The
method's adaptation to black hole perturbation theory emerged in the late
1980s, notably with the work of Majumdar and Panchapakesan
\cite{Majumdar:1989tzg}, who demonstrated its effectiveness for determining
the complex QNM frequencies of Schwarzschild black holes. More recently, the
method has been further refined and extended, incorporating convergence
acceleration techniques such as the Wynn algorithm and Borel summation, which
have enabled high-precision calculations for higher overtones and for
spacetimes in higher dimensions \cite{Benda:2025tni, Matyjasek:2024uwo, Matyjasek:2021xfg, Konoplya:2011qq}. In particular, Ref. \cite{Benda:2025tni} have demonstrated that the Hill
determinant approach, combined with double convergence acceleration and
Leaver-Nollert-Zhidenko tail approximations, achieves exceptional accuracy and
stability, even for modes with small real parts.

We start from the five-term recurrence. The condition for the existence of nontrivial solutions of the recurrence relation is given by
\begin{equation}
    \det\mathcal{H}=0,
\end{equation}
where $\mathcal{H}$ is the Hill matrix of width
\begin{equation}
    \mathcal{H=}\left(
\begin{array}
[c]{cccccc}%
\beta_{0} & \alpha_{0} & 0 & 0 & 0 & ...\\
\gamma_{1} & \beta_{1} & \alpha_{1} & 0 & 0 & ...\\
\delta_{2} & \gamma_{2} & \beta_{2} & \alpha_{2} & 0 & ...\\
\sigma_{3} & \delta_{3} & \gamma_{3} & \beta_{3} & \alpha_{3} & ...\\
0 & \sigma_{4} & \delta_{4} & \gamma_{4} & \beta_{4} & \ddots\\
\vdots & \vdots & \ddots & \ddots & \ddots & \ddots
\end{array}
\right)  .
\end{equation}

 We consider the determinants of the $N \equiv n\times n$ leading principal
submatrices $H_{n}$, whose main diagonal consists of the entries $\beta
_{0},...,\beta_{n-1}$. Alternatively, one may employ a simple formula for the construction of the determinants.  Denoting by $h_{n}$ the determinant of the $(n+1)\times (n+1)$ matrix, for $n \geq 4$ one obtains
\begin{align}
h_n = \beta_n h_{n-1} - \gamma_n \alpha_{n-1} h_{n-2} \nonumber
      + \delta_n \alpha_{n-1}\alpha_{n-2} h_{n-3} \\
      - \varepsilon_n \alpha_{n-1}\alpha_{n-2}\alpha_{n-3} h_{n-4}. \label{eq:recurrencehill}
\end{align}
These determinants define polynomial equations in
$\rho$, which can be systematically analyzed in increasing matrix order.
However, imposing the additional condition that the waves be purely outgoing
at spatial infinity, along with the requirement of stability, restricts the
allowed frequencies $\omega$ to those lying in the complex plane with negative
imaginary parts. These correspond precisely to the quasinormal modes. 
An important point of the Hill determinant method, in contrast to the continued
fraction approach Ref. \cite{Leaver:1985ax}, is that it remains applicable to
recurrence relations involving more than three terms. Consequently, due to the sparsity of the Hill matrices and the representation of the determinant in the form of Eq.~(\ref{eq:recurrencehill}), there is no practical need to reduce the original five-term recurrence relation to a tridiagonal form via Gaussian elimination.

Our strategy for applying Hill's method proceeds as follows. 
Initially, we employ \texttt{FindRoot} in \textit{Mathematica} for matrices of size $N=100$, 
using as an initial guess the value of the massless quasinormal mode available. 
The root obtained in this step is then used as the initial guess to compute the subsequent quasinormal modes. 
For example, for $l=0$, we take $\omega = 0.1105 - 0.10491\,i$ as the initial guess to determine the quasinormal mode with mass $m = 0.01$. 
Once this value is determined, it is used as the initial guess for $m = 0.02$, and so on.\footnote{We adopt steps of $m = 0.01$ to improve the stability of the method and the reliability of the initial guesses.
}

In the next section, we present Leaver's method, and in Sec. \ref{sec:results}, we compare the results obtained from both methods.

\section{The Leaver continued fraction method} \label{sec:leavermethod}
 Leaver's method relies on applying the Frobenius method for differential equations, expressing the solution of the perturbation equation as a power series around the event horizon. It can be shown that this series satisfies the quasinormal mode (QNM) boundary conditions only if a certain equation involving an infinite continued fraction is fulfilled. By solving this continued fraction equation with a root-finding algorithm, one can determine the QNM frequencies. 

 To apply Leaver's method in our case, it is first necessary to rewrite Eq.~(\ref{eq:RadialEq}) in terms of the tortoise coordinates, defined by
\begin{equation}
    dr^* = \frac{dr}{f(r)},
\end{equation}
so that the radial equation takes the form
\begin{equation}
\Phi''\left(  r^*\right)  
\left(\omega^2-V(r^*)\right) \Phi\left(  r^*\right)  =0,
\label{eq:RadialEqtortoise}%
\end{equation}
with the event horizon and spatial infinity now mapped to $r^* \to -\infty$ and $r^* \to +\infty$, respectively.  An appropriate ansatz in this case is \cite{Konoplya:2004wg}
\begin{equation}
\Phi\left(  \omega,r\right)  =\left(  \frac{r-1}{r}\right)  ^{\rho
}r^{-\nu+\frac{m^2}{2\nu}}e^{-\nu r}\sum\limits_{j=0}^{\infty}a_{j}\left(
\frac{r-1}{r}\right)  ^{j}, \label{eq:AnsatzMassivetortoise}%
\end{equation}
with $\rho=-i\omega$ and $\nu=-i\sqrt{-\rho^{2}-m^{2}}$. Inserting
Eq. (\ref{eq:AnsatzMassivetortoise}) into Eq. (\ref{eq:RadialEqtortoise}) we arrive at the following three-term recurrence relation:
\begin{align}
\alpha_{0}a_{1}+\beta_{0}a_{0}  &  =0,\label{eq:ThreeTermRecurrent}\\
\alpha_{j}a_{j+1}+\beta_{j}a_{j}+a_{j-1}\gamma_{j} &  =0\text{, \ \ \ }j>0,\nonumber
\end{align}
with coefficients explicitly given by
\begin{align}
\alpha_j &= (j + 1)(j + 2\rho + 1), \label{eq:CoeffThreeTerm} \\
\beta_j &= -\frac{1}{2\nu}(\rho+\nu)[2(\rho+\nu)^2+(2j+1)(\rho+3\nu)] \nonumber \\
&\quad -2j(j+1)-1 -l(l+1),  \\
\gamma_j &= \left( j+\frac{(\rho+\nu)^2}{2\nu} \right)^2. \nonumber \\
\end{align}
Once the three-term recurrence relation is obtained, the standard Leaver’s method can be applied 
to determine the QNMs as the roots of the algebraic equation \cite{Leaver:1985ax,Leaver:1990zz}
\begin{equation}
\beta_{0} 
- \frac{\alpha_{0}\gamma_{1}}{\beta_{1} -} 
  \frac{\alpha_{1}\gamma_{2}}{\beta_{2} -} 
  \frac{\alpha_{2}\gamma_{3}}{\beta_{3} -} 
  \cdots = 0.
\label{eq:CF2}
\end{equation}
To improve the convergence of the continued fraction on the right-hand side of Eq.~(\ref{eq:CF2}), we employ the technique developed by Nollert in Ref.\cite{Nollert:1993zz}. He demonstrated that the convergence of the continued-fraction method is enhanced if the sum is initiated with an appropriate estimate for the rest of the continued fraction, $R_N$, defined by  
\begin{equation}
R_N = \frac{\gamma_{N+1}}{\beta_{N+1} - \alpha_{N+1} R_{N+1}} ,
\label{eq:RNdef}
\end{equation}
where $\alpha_j$, $\beta_j$, and $\gamma_j$ are the recurrence coefficients.  

Assuming that the remainder can be expanded as an asymptotic series of the form
\begin{equation}
R_N = \sum_{k=0}^\infty C_k N^{-k/2} ,
\label{eq:RNasymp}
\end{equation}
the first few coefficients $C_k$ are
\begin{eqnarray}
    C_0 &=& -1,\\
    C_1 &=& \pm \sqrt{2i\omega},\\
    C_2 &=& \left( \frac{3}{4}+2i\omega\right).
\end{eqnarray}

Equation~(\ref{eq:CF2}) with Nollert improvement is solved numerically by employing the same strategy as in 
Sec.~\ref{sec:hilldeterminant}, using \texttt{FindRoot} in \textit{Mathematica} 
with $j=100$ and adopting the same procedure for generating initial guesses.

\section{Effective Potential and Critical Mass \label{sec:potentialandmass}}
\subsection{Roots and behavior near horzion}
The potential described by Eq. (\ref{eq:EffectivePot}) exhibits several important features. From a physical perspective, our focus lies on real and positive values of $r$. A straightforward analysis of Eq.(\ref{eq:EffectivePot}) reveals the following properties:
\begin{itemize}
    \item $r = 0$ corresponds to an essential singularity,
    \item $r = 1$ is the event horizon,
    \item $\lim_{r\to\infty} V(r) = m^2$.
\end{itemize}
The next step is to analyze the behavior of $V(r)$ near the horizon. To this end, we expand Eq.(\ref{eq:EffectivePot}) around $r = 1$, introducing a small parameter $\varepsilon > 0$, and obtain:
\begin{equation}
    V(1 \pm \varepsilon) \approx \pm (m^2+l(l+1)+1)\varepsilon.
\end{equation}
Hence, 
\begin{equation}
    V(1-\varepsilon) < 0, \ \ \ V(1+\varepsilon) > 0,
\end{equation} 
indicating that $V(r)$ increases near the horizon. Another important point is the existence (or absence) of roots outside the horizon. To investigate this, we rewrite the potential as
\begin{equation}
    V(r) = m^2 - \frac{m^2}{r} + \frac{l(l+1)}{r^2} - \frac{l(l+1)-1}{r^3} - \frac{1}{r^4},
    \label{eq:EffectivePotential2}
\end{equation}
and the roots, if they exist, correspond to the solutions of the equation
\begin{equation}
    m^2 r^3 + l(l+1)r + 1 = 0.
    \label{eq:CubicEq}
\end{equation}
Equation~(\ref{eq:CubicEq}) is a cubic equation, and in general, it presents three possible scenarios for its roots, depending on the value of the discriminant $\mathcal{D}$. These scenarios are
\begin{itemize}
    \item If $\mathcal{D}>0$, there is one real root and a pair of complex conjugate roots.
    \item If $\mathcal{D}<0$, all roots are real and distinct.
    \item If $\mathcal{D}=0$, all roots are real, with at least two equal.
\end{itemize}
In our case, the discriminant is given by
\begin{equation}
    \mathcal{D}= \left( \frac{l(l+1)}{3m^2} \right)^3 + \left( \frac{1}{2m^2} \right)^2,
    \label{eq:Discriminant}
\end{equation}
which is clearly positive. Therefore, Eq.~(\ref{eq:CubicEq}) has only one real root.
On the other hand, we have
\begin{equation}
    \lim_{r \to 1^+} V(r) > 0, \qquad \lim_{r \to \infty} V(r) = m^2 > 0.
\end{equation}
Therefore, we observe that for $r > 1$, the function $V(r)$ must either have two real roots greater than $1$ (i.e., it crosses the $x$-axis twice for $r > 1$) or no real roots at all (i.e., it never crosses the $x$-axis for $r > 1$). Combining this observation with the result above, we conclude that there are no roots for $r > 1$, and thus the potential is strictly positive in the region between the horizon and infinity.

\subsection{Critical masses}

The next step is to study the critical points of the potential. This is important because these points are associated with the critical masses, which in turn produce qualitative changes in the behavior of the quasinormal mode spectrum. By differentiating Eq.~(\ref{eq:EffectivePot}) by $r$ and setting the result to zero, we obtain
\begin{equation}
\frac{m^2}{r^2} - \frac{2l(l+1)}{r^3} + \frac{3[l(l+1)-1]}{r^4} + \frac{4}{r^5} = 0,
\label{eq:CubicEqDeriv}
\end{equation}
which leads to the following cubic equation:
\begin{equation}
m^2 r^3 - 2l(l+1)r^2 + 3[l(l+1)-1]r + 4 = 0.
\label{eq:P3_deriv}
\end{equation}
Before addressing the general case, let us consider the massless case. For $m = 0$, Eq.~(\ref{eq:P3_deriv}) becomes a quadratic equation:
\begin{equation}
    2l(l+1)r^2 - 3[l(l+1)-1]r - 4 = 0,
    \label{eq:QuadraticEqMassless}
\end{equation}
whose Vieta’s formulas give
\begin{equation}
    r_1 + r_2 = \frac{3[l(l+1)-1]}{2l(l+1)}, \qquad
r_1 r_2 = -\frac{2}{l(l+1)}.
\end{equation}
For positive \( l \), it is evident that one root is positive and the other negative. A more detailed analysis shows that the positive root is always greater than 1, indicating that the critical point lies outside the event horizon.\footnote{For \( l = 0 \), there is only one positive root given by \( r = \frac{4}{3} > 1 \), and thus also outside the horizon.}

Moreover, due to the structure of the potential with nonzero \( m \), this critical point corresponds to a maximum, forming a potential barrier. Consequently, the configuration consisting of the event horizon and the barrier behaves analogously to a resonant cavity, producing quasinormal modes.\footnote{The analogy is not perfect: in a typical resonant cavity (e.g., an optical cavity), there is usually a perfect reflector (a back mirror) and a partially reflective/transmissive region (a partially transparent mirror). In the black hole case, we have a perfect absorber (the horizon) and a partially transmissive barrier. Replacing the back mirror in the optical cavity with a perfect absorber (e.g., a black film) yields a much closer analogy to the black hole scenario. In both cases, perturbations generated by an external source excite modes that are absorbed by the perfect absorber and decay while escaping to infinity, i.e., quasinormal modes.}

We now consider the general case. Vieta’s formulas for Eq.~(\ref{eq:P3_deriv}) give:
\begin{align}
r_1 + r_2 + r_3 &= \frac{2l(l+1)}{m^2}, \label{eq:G1} \\
r_1r_2 + r_2r_3 + r_1r_3 &= \frac{3[l(l+1)-1]}{m^2}, \label{eq:G2} \\
r_1r_2r_3 &= -\frac{4}{m^2}. \label{eq:G3}
\end{align}

Since this is a cubic equation, we expect one of the following three scenarios:
\begin{itemize}
    \item All roots are real and distinct: the signs of Eqs.~(\ref{eq:G1}) and (\ref{eq:G3}) imply one root is negative and the other two are positive.
    \item All roots are real with at least two equal: again, the signs imply the repeated roots are positive and the remaining one is negative.
    \item One real root and two complex conjugates: Eq.~(\ref{eq:G3}) indicates that the real root is negative.
\end{itemize}
These three scenarios correspond to different signs of the discriminant $\mathcal{D}$. It is therefore useful to compute the discriminant of Eq.~(\ref{eq:P3_deriv}). Let us define $L \equiv l(l+1)$, so that Eq.~(\ref{eq:P3_deriv}) becomes
\begin{equation}
    m^2 r^3 - 2L r^2 + 3(L - 1)r + 4 = 0.
\end{equation}
Using the general formula for the discriminant of a cubic, we find:
\begin{align}
\mathcal{D}(m,L) &= \frac{1}{m^8} \Bigg[ 4m^4 + (L+1)(L^2 - 1)m^2 \nonumber \\
&\quad - \left( \frac{1}{3}L^2 + \frac{14}{27}L + \frac{1}{3} \right)L^2 \Bigg].
\label{eq:DmL}
\end{align}

The expression above depends on \( m \) and \( L \) and can be positive, negative, or zero. In fact, if we fix \( L \) and gradually increase \( m^2 \), we see that \( \mathcal{D}(m,L) \) transitions from negative (small \( m^2 \)) to positive (large \( m^2 \)). It is thus useful to solve \( \mathcal{D}(m,L) = 0 \). The critical values \( m^2_{\text{max}} \) that mark this transition are
\begin{equation}
m_{\text{max}}^2 = \frac{1}{8} \left[ \sqrt{(L^2 + \tfrac{2}{3}L + 1)^3} - (L+1)^2(L - 1) \right].
\label{eq:m_max}
\end{equation}

Based on Vieta’s formulas, the discriminant in Eq.~(\ref{eq:DmL}), and the value of \( m_{\text{max}} \), we can analyze how the potential changes with increasing \( m \). Let us examine three qualitatively distinct regimes, which affect the perturbation spectrum:\footnote{Although we define \( L \equiv l(l+1) \), all the results below remain valid even for \( L = 0 \).}

\begin{itemize}
    \item For small \( m^2 \), the discriminant \( \mathcal{D}(m,L) < 0 \), and the cubic has two positive roots and one negative root. From Eq.~(\ref{eq:G3}), one of the positive roots \( r_3 \) lies deep in the radial domain. The other root \( r_2 \), by continuity from the massless case, lies outside the horizon, and must be negative. Recalling that
\begin{equation}
       V(1) = 0, \ \lim_{r \to \infty} V(r) = m^2, \ V(r) > 0 \text{ for } r > 1,
\end{equation}
    we conclude that \( 1 < r_2 \ll r_3 \), where \( r_2 \) is a local maximum and \( r_3 \) a minimum. Thus, in addition to a potential barrier at \( r_2 \), we also have a potential well with minimum at \( r_3 \). In the limit \( m \to 0 \), we recover the massless case as \( r_3 \to \infty \).

    \item As \( m^2 \) increases, \( r_3 \) decreases and approaches \( r_2 \). Simultaneously, the asymptotic value of the potential increases. At a critical value, the asymptotic value equals the barrier peak. This occurs when the equation \( V(r) = m^2 \) has a single root. From Eq.~(\ref{eq:EffectivePot}), this yields:
 \begin{equation}
         m^2 r^3 - L r^2 + (L - 1)r + 1 = 0.
 \end{equation}
    This cubic has a single root only when its discriminant vanishes:
\begin{align}
27m^4 + 2(2L^3 + 3L^2 - 3L - 2)m^2 \\
- (L^2 + 2L + 1)L^2 &= 0. \nonumber
\end{align}
    Solving for \( m^2 \), we obtain the critical value
\begin{align}\label{eq:limmax}
m_{\text{lim}}^2 &= \frac{2}{27} \Big[ \sqrt{(L^2 + L + 1)^3} \nonumber \\
&\quad - \left(L^3 + \tfrac{3}{2}L^2 - \tfrac{3}{2}L - 1\right) \Big].
\end{align}
 For \( m < m_{\text{lim}} \), the barrier peak is higher than the asymptotic potential. For \( m > m_{\text{lim}} \), the situation is reversed. During the transition at \( m = m_{\text{lim}} \), the point \( r_2 \) becomes a local maximum instead of a global one.
    \item Numerical analysis shows that \( m_{\text{lim}} < m_{\text{max}} \). Therefore, in the interval \( m_{\text{lim}} < m < m_{\text{max}} \), the points \( r_2 \) and \( r_3 \) remain a maximum and minimum, respectively. When \( m = m_{\text{max}} \), the discriminant vanishes and \( r_2 = r_3 \), characterizing a point of inflection. For \( m > m_{\text{max}} \), the potential has no critical points outside the horizon and becomes strictly increasing.
\end{itemize}

The limit mass $m_{lim}$ given by Eq.~(\ref{eq:limmax}) has a form similar to that found in Ref.~\cite{Simone:1991wn}. In that work, the corresponding value is referred to as the maximum mass, defined as the largest mass for which the effective potential still exhibits a peak, and beyond this value, the peak disappears. According to the analysis of QNMs in Ref.~\cite{1985ApJ...291L..33S}, quasinormal modes can be interpreted as waves trapped by this peak. Thus, once the peak vanishes, the potential can no longer confine waves, and QNMs cease to exist. In contrast, in the present work, we introduce a different threshold, $m_{\text{lim}}$, defined as the mass at which the potential peak equals its asymptotic value. Thus, for $m > m_{\text{lim}}$, the potential still retains a peak, and we still have QMNs.

With the preceding discussion in mind, we present below a table containing several values of $m_{\text{lim}}$ and $m_{\text{max}}$ for different values of $l$. 
\begin{table}[ht!]
\centering
\begin{tabular}{|c|c|c|c|c|c|c|c|}
\hline
$l$ & 0 & 1 & 2 & 3 & 4 & 5 \\
\hline
$m_{\text{lim}}$ & 0.385 & 0.794 & 1.276 & 1.768 & 2.264 & 2.761 \\
\hline
$m_{\text{max}}$ & 0.500 & 0.931 & 1.480 & 2.047 & 2.618 & 3.192 \\
\hline
\end{tabular}
\caption{Values of the lower mass limit \( m_{\text{lim}} \) and the upper bound mass \( m_{\text{max}} \) for different values of the angular momentum number \( l \).}
\label{table:mass_bounds}
\end{table}

Furthermore, using the values of $m$ listed in Table~\ref{table:mass_bounds}, we plot the behavior of the effective potential for $l = 0$ and $l = 1$ with various values of $m$.
\begin{figure*}[ht!]
\centering
\begin{minipage}{0.48\textwidth}
  \centering
  \includegraphics[width=\linewidth]{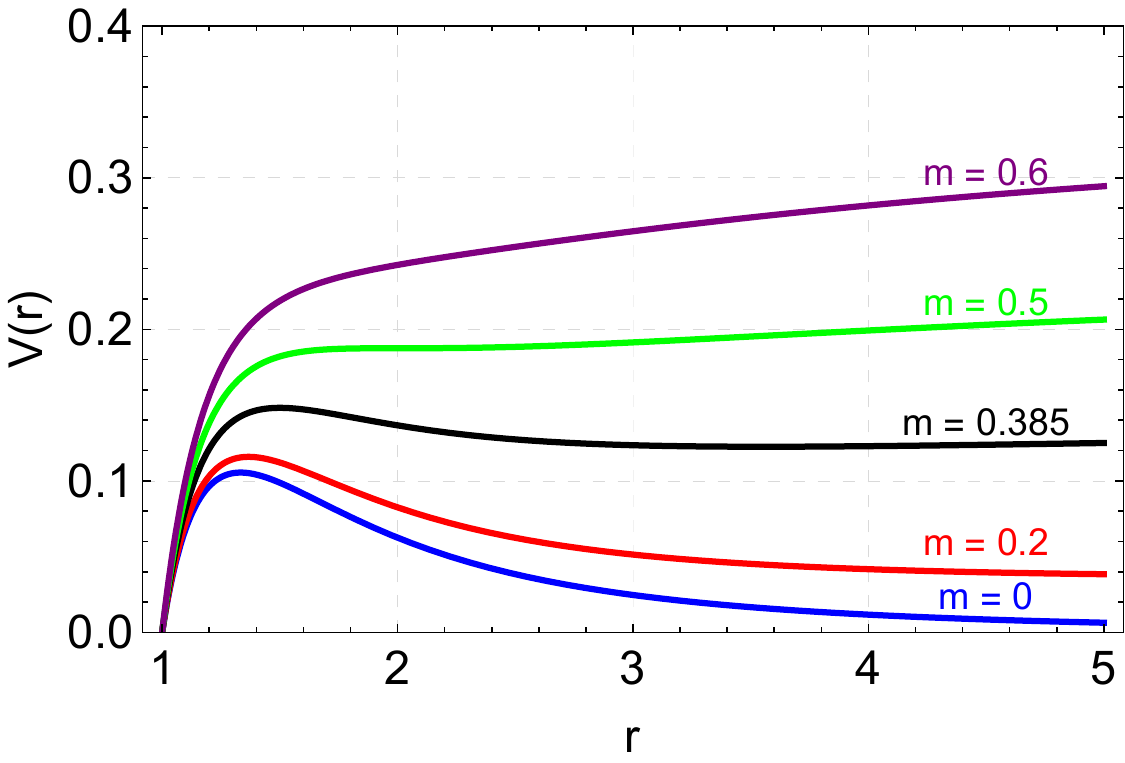}
\end{minipage}
\hfill
\begin{minipage}{0.48\textwidth}
  \centering
  \includegraphics[width=\linewidth]{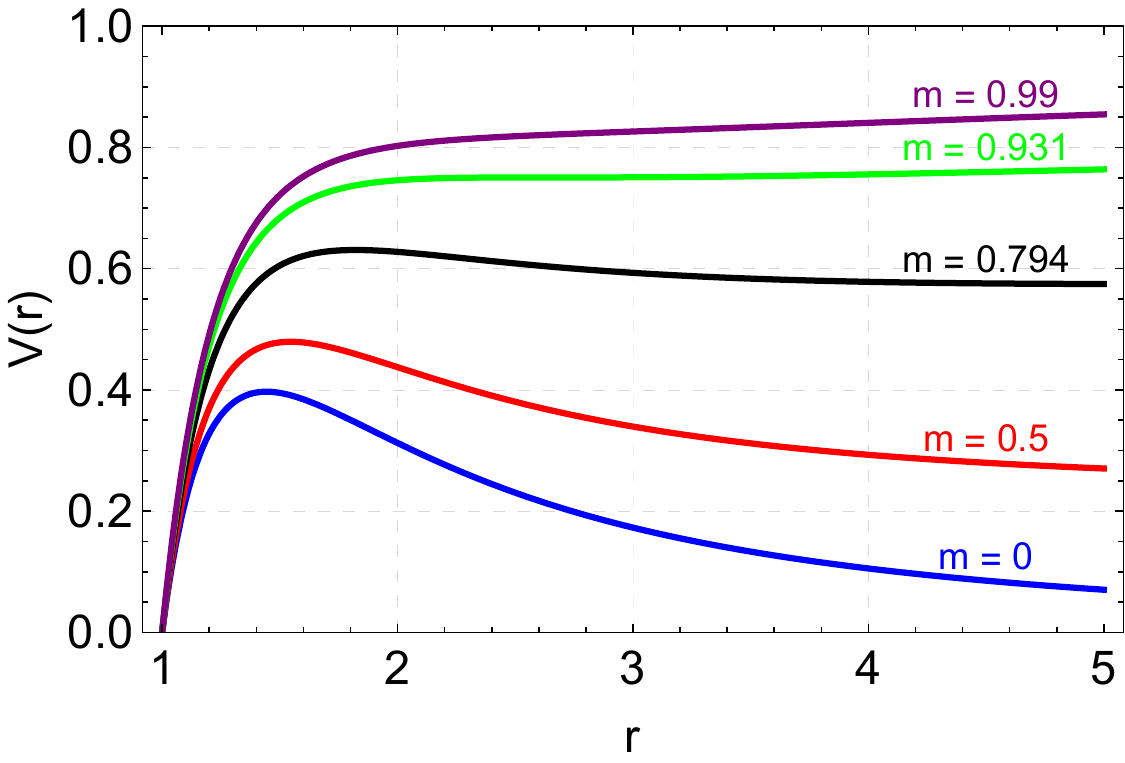}
\end{minipage}
\caption{The effective potential $V$ as a function of $r$ for $l = 0$ (left panel) and $l=1$ (right panel), for different values of mass $m$.}
\label{fig:EffectivePotential}
\end{figure*}

Figure \ref{fig:EffectivePotential} shows the behavior of the effective potential for $l = 0$ and $l = 1$, respectively. As discussed earlier in this section, increasing the mass $m$ causes the potential peak to approach the asymptotic value. At $m = m_{\text{lim}}$, the global maximum becomes clearly a local maximum and transitions into an inflection point when the mass reaches $m = m_{\text{max}}$. For $m > m_{\text{max}}$, the potential peak disappears entirely.

\section{RESULTS \label{sec:results}}
\subsection{Massless limit}
As the first part of our analysis, we examine the massless limit and compute the corresponding QNMs. This is achieved by solving the five-term recurrence relation given in Eq.(\ref{eq:FiveTermRecurrent}), with the coefficients defined in Eq.(\ref{eq:CoeffFiveTermMassless}), using the method outlined in Sec. \ref{sec:hilldeterminant}. We compare our results with those obtained via Leaver's continued fraction method, as presented in Sec. \ref{sec:leavermethod}. The computed values are summarized in the Table \ref{table1}.
\begin{table}[ht!]
\centering
\begin{tabular}{|c|c|c|c|c|}
\hline
$(l, n)$ & $\omega_{\text{Hill}}$ & $\omega_{\text{Leaver}}$ & $\epsilon_{\text{Re}}$ & $\epsilon_{\text{Im}}$ \\
\hline
$(0,0)$ & $0.1105 - 0.1049i$ & $0.1105 - 0.1049i$ & $0.00\%$ & $0.00\%$ \\
\hline
$(0,1)$ & $0.0859 - 0.3478 i$ & $0.0861 - 0.3481i$ & $0.23\%$ & $0.09\%$ \\
\hline
$(1,0)$ & $0.2929 - 0.0977i$ & $0.2929 - 0.0977i$ & $0.00\%$ & $0.00\%$ \\
\hline
$(1,1)$ & $0.2644 - 0.3063i$ & $0.2645 - 0.3063i$ & $0.04\%$ & $0.00\%$ \\
\hline
$(1,2)$ & $0.2295 - 0.5401i$ & $0.2295 - 0.5401i$ & $0.00\%$ & $0.00\%$ \\
\hline
$(2,0)$ & $0.4836 - 0.0968 i$ & $0.4836 - 0.0968i$ & $0.00\%$ & $0.00\%$ \\
\hline
$(2,1)$ & $0.4639 - 0.2956i$ & $0.4639 - 0.2956i$ & $0.00\%$ & $0.00\%$ \\
\hline
$(2,2)$ & $0.4305 - 0.5086i$ & $0.4305 - 0.5086i$ & $0.00\%$ & $0.00\%$ \\
\hline
\end{tabular}
\caption{Comparison between quasinormal mode frequencies computed using
the Hill determinant method ($\omega_{Hill}$) and the Leaver method ($\omega_{Leaver}$). The last two columns are the relative errors in the real ($\epsilon_{\text{Re}}$) and imaginary ($\epsilon_{\text{Im}}$) parts of the QNM frequencies computed using the Hill determinant and Leaver methods.}
\label{table1}
\end{table}

Table \ref{table1} compares quasinormal mode frequencies computed using
the Hill determinant method with those obtained via Leaver's method. The two approaches exhibit excellent
agreement, with relative errors consistently small and many cases
showing differences within numerical precision. This strong correspondence
validates the accuracy of the Hill determinant method for calculating
quasinormal modes in Schwarzschild spacetime and confirms its ability to
reproduce the well-established spectrum from Leaver's continued fraction
approach. Furthermore, despite our formulation involving a five-term
recurrence relation, both methods yield the same quasinormal mode spectrum consistent with the findings of Ref.~\cite{Iyer:1986np}.

\begin{figure*}[ht!]
\centering
\begin{minipage}{0.48\textwidth}
  \centering
  \includegraphics[width=\linewidth]{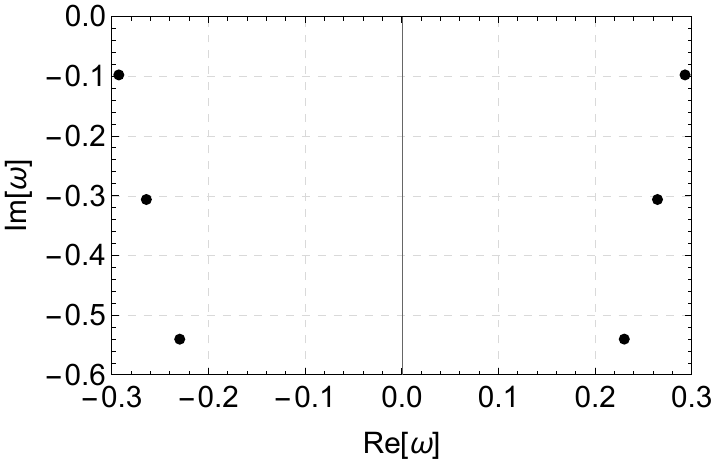}
\end{minipage}
\hfill
\begin{minipage}{0.48\textwidth}
  \centering
  \includegraphics[width=\linewidth]{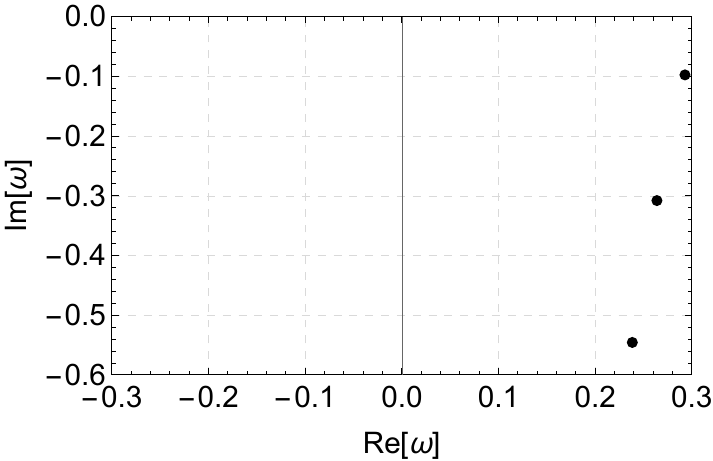}
\end{minipage}
\caption{Quasifrequency spectra for the fundamental mode with $l=1$. Left panel: massless case ($m=0$), showing symmetric QNMs with respect to $\text{Re}(\omega)$. Right panel: massive case ($m = 10^{-5}$), where the symmetry is broken.}
\label{fig:masslesslimit}
\end{figure*}

Another relevant aspect to highlight is the difference in the distribution of quasinormal modes between the massive and massless cases. Looking at the discussion made in Sec.\ref{sec:massivescalar}, we see that in the massless limit ($m=0$), only QNMs are present. These modes exhibit a characteristic spectral symmetry, as shown in 
Figure \ref{fig:masslesslimit} (left panel). This symmetry was discussed previously in Ref. \cite{Iyer:1986nq}. In contrast, when a small mass is introduced, this symmetry is broken and quasibound states emerge, as illustrated in the right panel of Figure~\ref{fig:masslesslimit}.

Figure \ref{fig:masslesslimit} (right panel) illustrates the effect of introducing mass into the scalar field. In this case, the symmetry present in the massless regime is broken. It is important to highlight that, despite the presence of a small mass ($m = 10^{-5}$), the values of the quasinormal modes (Re$(\omega)>0$) remain essentially unchanged when compared to the massless case. The key difference lies in the breaking of the spectral symmetry.

Having studied the massless case and confirmed the accuracy of the Hill determinant method through comparison with Leaver’s method, 
we now turn to the massive case. In the next subsection, we compute the corresponding quasinormal modes using both approaches and 
carry out a detailed comparison.

\subsection{Quasinormal modes}

In the previous subsection, we showed that the Hill determinant method and Leaver’s method yield comparable results in the massless case. 
Before analyzing their performance in the massive regime, we first determine the quasinormal modes for a fixed mass using the Hill determinant method. 
Subsequently, we compare the two methods across different values of $m$.
\begin{table*}[ht!]
\centering
\begin{tabular}{|c|c|c|c|}
\hline
$l$ & $\omega\ (n=0)$ & $\omega\ (n=1)$ & $\omega\ (n=2)$  \\
\hline
0 & $0.110988 - 0.102842 i$ & $0.0855785 - 0.348144 i$ & $-$ \\
\hline
1 & $0.294054 - 0.096988 i$ & $ 0.264512 - 0.305404 i$ & $0.229261 - 0.539697  i$ \\
\hline
2 & $0.484433 - 0.0964882 i$ & $0.464256 - 0.29502 i$ & $0.430575 - 0.508034 I i$  \\
\hline
3 & $0.675956 - 0.0963572 i$ & $0.633826 - 0.495577i$ & $0.598794 - 0.710845 i$ \\
\hline
4 & $0.867883 - 0.0963044 i$ & $0.833947 - 0.490004 i$ & $0.423585 - 1.59423 i$ \\
\hline
5 & $1.06000  - 0.0962779 i$ & $1.05038 - 0.289989 i$ & $0.973049 - 0.899212 i$ \\
\hline
6 & $1.25222 - 0.0962628 i$ & $1.22809 - 0.48542 i$ & $1.17615 - 0.890251 i$ \\
\hline
\end{tabular}
\caption{Quasinormal modes computed using
the Hill determinant method for $m=0.1$ in Schwarzschild spacetime.
All frequencies are given in units where $2M=1$.}
\label{table2}
\end{table*}

Table \ref{table2} presents the quasinormal modes for a massive
scalar field with $m=0.1$ as a function of the angular momentum number $l$ and
overtone number $n$. Note that, for each fixed $l$, the real part of the
fundamental frequency ($n=0$) increases monotonically with $l$, reflecting the higher
oscillation frequencies associated with modes of greater angular momentum.
This is accompanied by a gradual decrease in the magnitude of the imaginary part, which indicates that higher $l$ modes are longer-lived. This result is consistent with previous studies on massive scalar perturbations \cite{Ma:2006by,
Zinhailo:2024jzt, Konoplya:2005et}.

The quasinormal spectrum of a massive field exhibits distinctive and nontrivial features, first reported in Ref.~\cite{1985ApJ...291L..33S}. In contrast to the comparatively simple spectra of massless fields, the presence of mass introduces characteristic structures in the frequency domain. As the field mass parameter \( m \) increases, the damping rate of the fundamental (least damped) mode decreases steadily, approaching zero at a critical value of \( m \). Beyond this threshold, the fundamental mode disappears from the spectrum, and the first overtone assumes dominance. This pattern repeats for progressively larger \( m \), with successive overtones undergoing the same transition. Consequently, the spectrum of massive fields may contain quasinormal modes with arbitrarily small damping, corresponding to extremely long-lived oscillations that closely resemble standing waves.

As the overtone number $n$ increases for a given $l$, the real part of the
frequency decreases, while the magnitude of the imaginary part increases. This
behavior signifies that overtones oscillate more slowly and decay more rapidly
than the fundamental mode, in line with the general properties of quasinormal modes in
black hole spacetime.

The presence of the scalar field mass manifests as an overall shift in the
quasinormal modes spectrum: the real part of the frequencies is slightly increased compared
to the massless case, while the imaginary part is reduced in magnitude,
leading to longer-lived perturbations. This tendency, observed across all $l$ and
$n$, is in agreement with theoretical expectations and previous numerical
results \cite{Becar:2023zbl, Lu:2023par}. The slower decay of
massive field perturbations is particularly relevant for gravitational wave
phenomenology, as it implies that massive scalar fields can produce late-time
signals with potentially observable imprints \cite{Zinhailo:2024jzt}.

\begin{figure*}[ht!]
\centering
\begin{minipage}{0.48\textwidth}
  \centering
  \includegraphics[width=\linewidth]{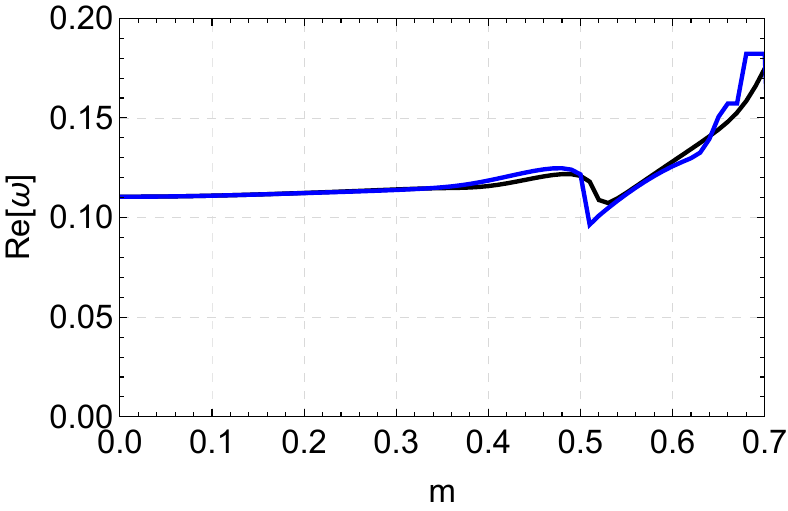}
\end{minipage}
\hfill
\begin{minipage}{0.48\textwidth}
  \centering
  \includegraphics[width=\linewidth]{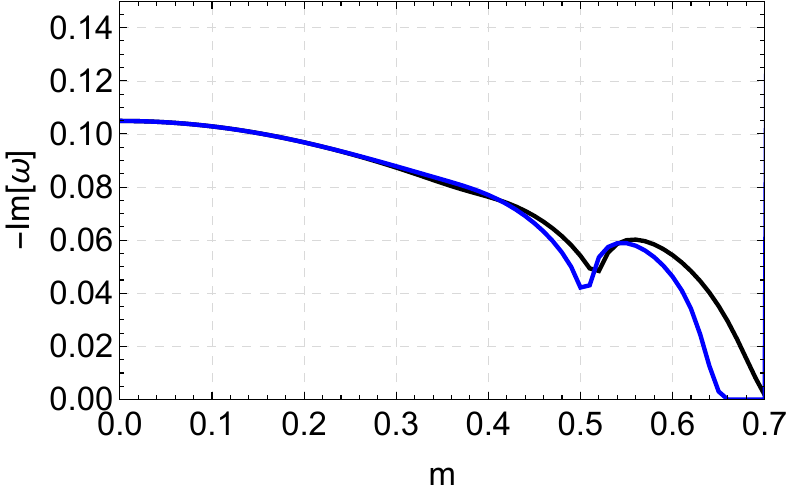}
\end{minipage}
\caption{Fundamental mode for $l=0$ QNMs of the scalar field as a function of the field mass, computed using the Leaver method (black curve) and the Hill determinant method (blue curve). Left panel: real part of the frequency. Right panel: imaginary part of the frequency.
}
\label{fig:model0qnm}
\end{figure*}

We now analyze the behavior of QNMs as the scalar field mass approaches the critical values $m_{\rm lim}$ and $m_{\rm max}$. We begin with the case $l=0$, for which the critical masses are $m_{\rm lim}=0.385$ and $m_{\rm max}=0.5$. The evolution of the fundamental mode with increasing mass is illustrated in Figure~\ref{fig:model0qnm}. The real part of the QNM frequency grows with the mass, while the imaginary part decreases and approaches zero at a certain mass value. A notable feature is the qualitative change in behavior observed at $m=0.5$: the real part exhibits a drop followed by a less smooth rise, while the imaginary part initially increases, then decreases, vanishing close to $m\simeq0.7$. This change occurs when the effective potential ceases to have a distinct peak, which, according to the interpretation of QNMs described in Ref.~\cite{1985ApJ...291L..33S}, would indicate the nonexistence of QNMs for masses beyond $m_{\rm max}$. Nevertheless, it is possible to obtain numerical frequency values beyond this mass, including cases where the imaginary part vanishes. Modes with vanishing imaginary parts are identified as long-lived modes, corresponding to perturbations with arbitrarily long lifetimes, and have been extensively studied in the literature~\cite{Ohashi:2004wr,Zinhailo:2024jzt,Konoplya:2017tvu,Konoplya:2005hr,Konoplya:2004wg}.

\begin{figure*}[ht!]
\centering
\begin{minipage}{0.48\textwidth}
  \centering
  \includegraphics[width=\linewidth]{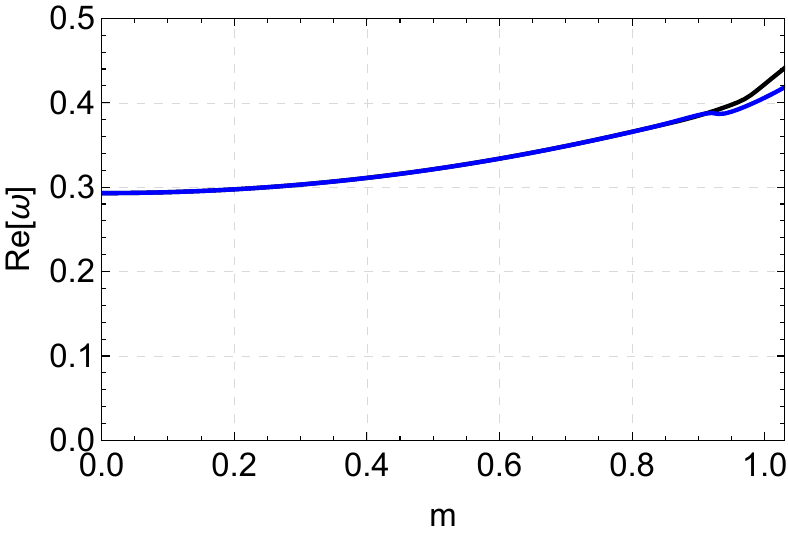}
\end{minipage}
\hfill
\begin{minipage}{0.48\textwidth}
  \centering
  \includegraphics[width=\linewidth]{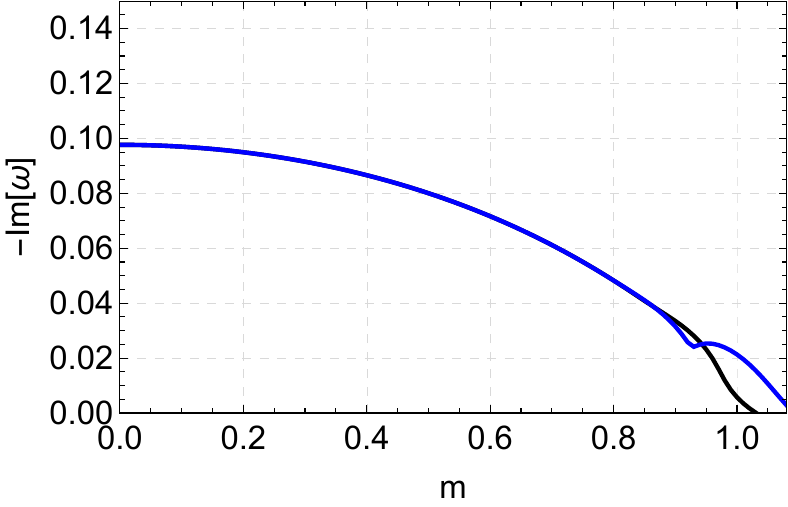}
\end{minipage}
\caption{Fundamental mode for $l=1$ QNMs of the scalar field as a function of the field mass, computed using the Leaver method (black curve) and the Hill determinant method (blue curve). Left panel: real part of the frequency. Right panel: imaginary part of the frequency.
}
\label{fig:model1qnm}
\end{figure*}

For $l=1$, the critical values are $m_{\rm lim}=0.794$ and $m_{\rm max}=0.931$. As shown in Figure~\ref{fig:model1qnm}, the real part of the QNM frequency again increases with the mass, while the imaginary part decreases as the mass approaches $m_{\rm max}$. A change in behavior seems to occur at $m = m_{\rm max}$, although the discrepancy between the Leaver and Hill methods makes it unclear whether this feature is real. Beyond this point, the imaginary part continues to decrease and eventually approaches zero near $m\simeq1.0$, indicating the emergence of a long-lived mode.

\begin{figure*}[ht!]
\centering
\begin{minipage}{0.48\textwidth}
  \centering
  \includegraphics[width=\linewidth]{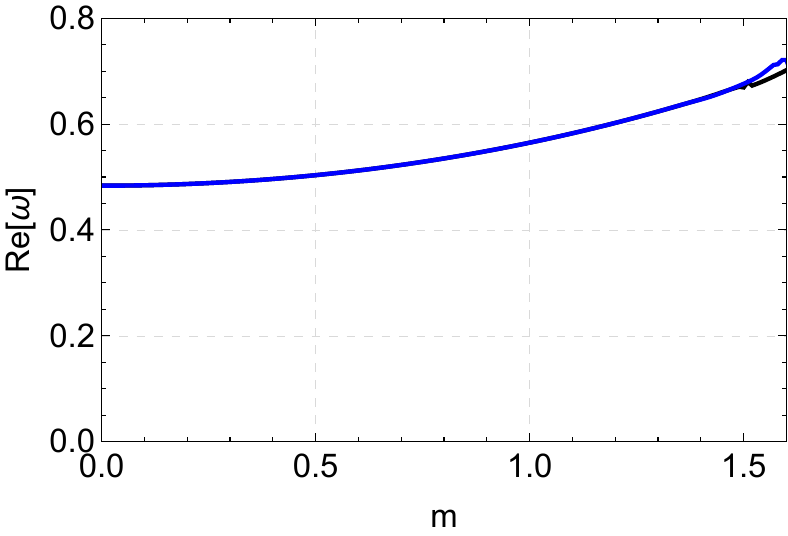}
\end{minipage}
\hfill
\begin{minipage}{0.48\textwidth}
  \centering
  \includegraphics[width=\linewidth]{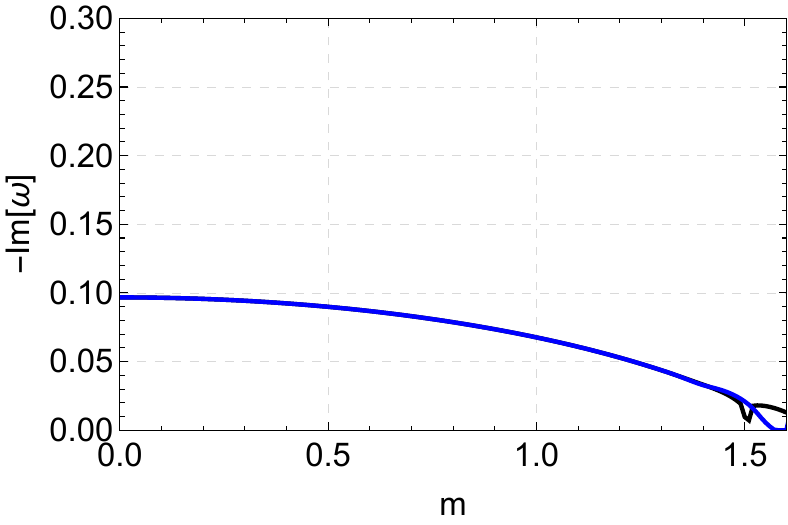}
\end{minipage}
\caption{Fundamental mode for $l=2$ QNMs of the scalar field as a function of the field mass, computed using the Leaver method (black curve) and the Hill determinant method (blue curve). Left panel: real part of the frequency. Right panel: imaginary part of the frequency.
}
\label{fig:model2qnm}
\end{figure*}

As shown in Figure~\ref{fig:model2qnm}, a similar trend is observed for $l=2$, with $m_{\rm lim}=1.276$ and $m_{\rm max}=1.480$. Nevertheless, long-lived modes still appear beyond $m_{\rm max}$, with the imaginary part vanishing near $m\simeq1.6$. This indicates that the emergence of long-lived modes occurs at different values of $l$, although the precise mass at which they appear shifts to higher values as $l$ increases.

In summary, for the $l=0$ mode, the real and imaginary parts of the QNM frequencies exhibit a clear qualitative transition near $m_{\rm max}$, which corresponds to the disappearance of the peak in the effective potential. A similar, though less evident, behavior is suggested for the $l=1$ and $l=2$ modes. This transition may indicate the nonexistence of physical QNMs for $m>m_{\rm max}$. Despite this, both numerical methods continue to yield frequency values beyond $m_{\rm max}$, consistently indicating the emergence of long-lived modes at a mass $m_{zd}$, referred to as the zero-damping mass. At this point, the imaginary part of the frequency vanishes, and the mode becomes effectively nondecaying. While the exact value of $m_{zd}$ differs slightly between the two methods, its existence is robust. By contrast, $m_{\rm lim}$ has no apparent effect on the qualitative behavior of the QNM spectrum. We therefore identify three critical mass scales: $m_{\rm max}$, which governs the qualitative transition in both the real and imaginary parts of the frequencies; $m_{\rm lim}$, which appears to be irrelevant for the spectral dynamics within the explored parameter space; and $m_{zd}$, which is not associated with the effective potential but rather signals the onset of long-lived modes.

Finally, regarding the comparison between the two numerical schemes in the massive case, we find that both methods yield consistent results up to field masses close to $m_{\text{max}}$. Beyond this value, discrepancies begin to appear in both the real and imaginary parts of the quasinormal frequencies, the most significant being the precise determination of the critical mass $m_{zd}$. Notably, this difference tends to diminish as the multipole number $l$ increases. We have also verified the numerical stability of both approaches with respect to the matrix size $N$ and the number of terms $j$, considering values $N=j=100$, $500$, and $1000$. In all cases, the results remain unchanged, confirming the high stability and reliability of both methods.

This slight divergence between the two approaches originates from their distinct mathematical formulations. In Leaver’s continued-fraction method, the quasinormal frequencies are obtained from the convergence condition of an infinite series whose behavior depends sensitively on the asymptotic structure of the recurrence coefficients. In contrast, the Hill-determinant method relies on truncating a finite matrix and imposing the vanishing of its determinant, thereby avoiding explicit convergence issues inherent to infinite series. A well-known case where the continued-fraction approach may fail is that of purely imaginary quasinormal frequencies, for which the recurrence relation becomes numerically unstable and convergence deteriorates \cite{Cook:2016ngj, Fortuna:2020obg}. In this regard, the Hill-determinant technique generally exhibits fewer convergence pathologies and often remains stable even in such extreme regimes. Nevertheless, this observation alone is not sufficient to determine which of the two methods performs better for the present problem. Despite this, qualitatively, both methods indicate the existence of $m_{zd}$, demonstrating that its presence is a robust feature of the spectrum.

\section{Conclusions}
\label{sec:conclusions}

In this work, we have presented a comprehensive analysis of the QNMs of a massive scalar field in Schwarzschild spacetime, employing two complementary numerical techniques: the Hill-determinant method and Leaver’s continued-fraction method. Both approaches proved robust and yielded consistent spectra across a wide range of field masses and angular momenta.

A central result of our study is the identification of three critical mass thresholds, $m_{\rm lim}$, $m_{\rm max}$, and $m_{zd}$, which govern qualitative changes in the QNM spectrum. As the scalar field mass approaches $m_{\rm max}$, the effective potential loses its characteristic peak, and the frequencies undergo a marked transition in behavior. Beyond this threshold, following the classical interpretation of Ref.~\cite{1985ApJ...291L..33S}, physical QNMs no longer exist. Nevertheless, our numerical analysis showed that there are solutions that evolve into long-lived, or zero-damping, states at a distinct value of the mass, denoted $m_{zd}$, where the imaginary part of the frequency vanishes and the mode becomes essentially nondecaying. Although the precise value of $m_{zd}$ shows minor discrepancies between the two methods, its presence is robust across all multipoles and shifts to larger values with increasing $l$.

We further established that the lower threshold $m_{\rm lim}$ exerts negligible influence on the qualitative features of the spectrum within the parameter space investigated. The emergence of long-lived modes at $m_{zd}$ may have potential observational consequences, particularly in the context of gravitational wave signatures of massive fields around black holes.

In summary, our study provides a detailed numerical characterization of massive scalar QNMs, clarifies the role of mass thresholds in shaping their spectra, and demonstrates the complementary strengths of the Hill-determinant and continued-fraction methods. Future extensions of this framework could include the analysis of rotating or charged black holes, as well as possible connections with quasibound states.

\section*{ACKNOWLEDGMENTS}

M.F.S.A. thanks \textit{Fundação de Amparo à Pesquisa e Inovação do Espírito Santo} (FAPES, Brazil) for support. L.G.M. thanks \textit{Conselho Nacional de Desenvolvimento Científico e Tecnológico} (CNPq, Brazil) for partial financial support—Grant: 307901/2022-0. B.P.P. thanks \textit{Coordenação de Aperfeiçoamento de Pessoal de Nível Superior} (CAPES) for financial support

\section*{DATA AVAILABILITY}
The data that support the findings of this article are openly available \cite{alves_2025_17580446}.

\bibliographystyle{apsrev4-1}
\bibliography{main}
\end{document}